\documentstyle[epsf,epsfig,eqsecnum,aps,amssymb,multicol]{revtex}

\begin{document}
\draft \preprint{HEP/123-qed}
\title{Dynamic ordering of driven vortex matter in the peak effect regime of amorphous MoGe films and 2H-NbSe$_2$ crystals}
\author{N. Kokubo$^1$ \cite{byline}, T. Asada$^1$, K. Kadowaki$^1$, K. Takita$^1$, T. G. Sorop$^2$ and P. H. Kes$^2$\\
}

\address{$^1$Institute of Materials Science, University of Tsukuba, 1-1-1, Tennoudai, Tsukuba, Ibaraki 305-8573, Japan\\
$^2$Kamerlingh Onnes Laboratory, Leiden University,P.O.Box 9504, 2300 RA Leiden, The Netherlands.\\}

\date{\today}
\maketitle
\begin{abstract}
Dynamic ordering of driven vortex matter has been investigated in the peak effect regime of both amorphous MoGe
films and 2H-NbSe$_2$ crystals by mode locking (ML) and dc transport measurements. ML features allow us to trace
how the shear rigidity of driven vortices evolves with the average velocity. Determining the onset of ML
resonance in different magnetic fields/temperatures, we find that the dynamic ordering frequency (velocity)
exhibits a striking divergence in the higher part of the peak effect regime. Interestingly, this phenomenon is
accompanied by a pronounced peak of dynamic critical current. Mapping out field-temperature phase diagrams, we
find that divergent points follow well the thermodynamic melting curve of the ideal vortex lattice over wide
field/temperature ranges. These findings provide a link between the dynamic and static melting phenomena which
can be distinguished from the disorder induced peak effect.
\end{abstract}

\pacs{PACS numbers: {74.25.Qt}, {74.25.Sv}, {74.25.Fy} }
\begin{multicols}{2} \narrowtext
\renewcommand{\theequation}{\arabic{equation}}

\section*{1. Introduction}
Vortex matter in type II superconductors has been recognized as an ideal system for studying collective
transports of periodic manifolds driven over the pinning environments
\cite{BhattHiggins1993}\cite{HigginsBhatt1996}. In particular, much recent attention has focused on the
issue of dynamic ordering, separating elasticity dominated, coherent lattice flow at large velocity,
from plastically deformed, incoherent defective flow at small velocity. Theoretical considerations for
this phenomenon were first proposed by Koshelev and Vinokur by introducing a concept of shaking
temperature, which characterizes fluctuating motion of vortices due to the interaction with disorder
quenched in a host material \cite{KosheleVinokurPRL1994}. Assuming the shaking temperature to decay as
$k_BT_{sh}=\Gamma/v$ on increasing the average velocity $v$, they proposed that the ordering
(crystallization) velocity diverges as $v_c \propto \Gamma/k_B(T_M-T)$ with a disorder parameter
$\Gamma$ on approaching the thermal melting temperature $T_M$ from below. Later, more sophisticated
theoretical \cite{DoussalGiamarchPRB1998}\cite{BalentsPRB1998}\cite{ScheidlPRE1998}\cite{ScheidlPRB1998}
and numerical studies \cite{FaleskiPRB1996}\cite{KoltonPRL1999}\cite{OlsonPRL1998}\cite{FangohrPRB2001}
\cite{ChandranPRB2003} \cite{GotchevaPRB2005} have suggested various coherent flow states with different
positional and orientational order like moving glass state and smectic flow state, and some of them have
proposed multiple transitions in ordering process.

While experimental evidences for the ordering were reported in neutron diffraction and Bitter decoration studies
\cite{Shelten1975}\cite{YaronNature1995}\cite{MarchevskyPRL1997}\cite{PardolNature1998}, investigations of the
divergence at the ordering point in dc transport studies were only based on the assumption that a S-shaped
anomaly in dc current-voltage characteristics marks current-induced dynamic ordering of driven vortices. In
NbSe$_2$ crystals \cite{BhattHiggins1993}\cite{HigginsBhatt1996}, the inflection anomaly is observed in the peak
effect regime, where the critical current shows a peak in the vicinity of the second critical field $H_{c2}$.
Meanwhile in amorphous ($\alpha$-)MoGe thin films \cite{HellerqvistPRL}, instead of the peak effect, the anomaly
is accompanied by a peak in the "dynamic" critical current, which characterizes a change of dynamic friction of
vortices in the flux-flow state. These studies claimed the divergence of the ordering current around these peak
behaviors and suggested the relevance of the thermal melting to the peaks of the static or dynamic pinning
force. However, a study of scanning Hall microscopy has shown that geometrical inhomogeneity in sample edges
leads to a macroscopic coexistence of ordered and disordered flow states and their dynamic evolutions with
transport current may result in similar S-shaped current-voltage characteristics \cite{MarchevskyNature2001}
\cite{PaltielPRB2002}. Thus, it has remained controversial if the S anomaly marks the dynamic order of driven
vortices.


Recently, mode locking (ML) has proved to be a powerful experimental technique for studying the dynamic ordering
of driven vortex matter. ML is a dynamic resonance between internal lattice modes of a driven lattice and an ac
drive superimposed on top of dc drive
\cite{FioryPRL1971}\cite{MartinoliSolidStateCom1975}\cite{LookPRB1999}\cite{KKBPRL2002}\cite{RutPRL2003}\cite{KKBPRB2004}\cite{KKBPRL2005}.
This technique directly probes the shear rigidity of driven vortices as a function of the velocity and allows us
to find the dynamic ordering (or melting) point from the onset (or disappearance) of ML resonance. As
demonstrated in a recent ML study on NbSe$_2$ crystals, the ordering velocity exhibits a steep increase in the
peak effect regime and this defines unambiguously the thermal melting point for driven vortex lattice
\cite{KKBPRL2005}.

To elucidate the relevance of thermal melting to the peak behaviors of the critical current and the dynamic
critical current, we will present in this article a systematic study of ML and dc transport measurements on both
$\alpha$-MoGe films and NbSe$_2$ crystals over wide magnetic field/temperature ranges. We will show how the
ordering frequency (velocity) grows with field and diverges in the peak effect regime.  We will map out
field-temperature phase diagrams and show how the divergent field, together with the peak fields of the critical
current and the dynamic critical current, grow with lowering temperature. We will discuss influences of quenched
disorder and thermal fluctuations on the onset of ML resonance and compare the divergent points to the
thermodynamic melting curve of the ideal vortex lattice proposed by a quantitative Ginzburg-Landau (GL) model
\cite{LiRosenPRB2002}. A brief comment on the peak behavior of the dynamic critical current will be made.

\section*{2. Experimental}
Amorphous Mo$_{1-x}$Ge$_x$ films ($x = 0.22\pm 0.01\%$) were sputtered on silicon substrates mounted on a
rotating copper stage held at room temperature by water cooling. The films were structured into a Hall-bar shape
by a liftoff technique. After partly deposition of silver on top of the films, thin gold wires were connected to
the films by using small indium pieces. We used two films (MG$\sharp$1 and MG$\sharp$2) with nearly identical
geometry (1.2 mm in length $l$, 0.3 mm in width $w$, and 0.33 $\mu$m in thickness $d$) and material parameters
of superconducting transition temperature $T_c$, normal resistivity $\rho_n$ and a slope of the second critical
field near $T_c$ $S(=\mu_0dH_{c2}/dT\mid _{T_c})$, which are given in table 1. We determined $T_c$ from
resistive transition by Aslamazov-Larkin fluctuation theory \cite{AL}. $H_{c2}$ for magnetic field perpendicular
to the films was determined from the lowest-landau-level (LLL) scaling analysis of fluctuation conductivity
\cite{TheuPRB1997}\cite{KKBprb2001}. The average composition for the films was obtained by the electron probe
micro analyzer.

We also used pure 2H-NbSe$_2$ single crystals grown by an iodine vapor transport method \cite{Takita1985}. The
thinnest platelets with optically flat surfaces on both sides were cut in a Bar-shape. We glued these on silicon
substrates with negative resist and cleaved them without observable surface damage. After deposition of silver
partly on crystal surface, thin gold wires were connected to voltage and current contacts by silver past and
indium solder, respectively. The geometry and material parameters for two cleaved crystals (NS$\sharp$1 and
NS$\sharp$2) used in this study were also summarized in table 1.  $T_c$ was determined by the mid-point of the
resistive transition, which has a transition width of about 50 mK between 10$\%$ and 90$\%$ of the normal state
resistance at 8 K, just above $T_c$.  Residual resistance ratio defined by $R$(295 K)/$R$(8 K) is about 30. We
applied magnetic field perpendicular to the ab plane of the crystals. $H_{c2}$ was determined from the
intersection between the linear extrapolation of flow resistance and the normal resistance \cite{SurfaceSuper}.
The thickness of the cleaved crystals was estimated from the room temperature resistance by $d=(l/w)\rho_n(295
\textrm{K})/R(295 \textrm{K})$ using the room temperature resistivity $\rho_n (295 K) = 1.2 \times 10^{-6}
\Omega$m determined before \cite{KKBPRL2005}.

We used Oxford cryostat with a VTI inset and a 17/19 superconducting magnet. We glued a sample on a copper stage
at the bottom of our measurement insert, and measured temperature by using a calibrated Cernox sensor attached
to the stage. For measurements below 4.2 K the sample was immersed in liquid of $^4$He introduced in the VTI
insert. We regulated the vapor pressure of the liquid by using a throttle valve, a capacitance manometer, a PID
controller and a rotary pump. Above 4.2 K, we used a temperature controller (Cryocon 62), which regulates
temperature of the vaporizer at the bottom of the VTI insert and that of the sample stage of the measurement
insert. We employed four-terminal pair method with a precision LCR meter (HP 4285A/001) and nanovoltmeters
(Keithley 2182).

The sample was wired with four-coaxial cables of which guards are connected. For a precise measurement of the ac
current in the frequency range of 0.1-30 MHz, we calibrated the coaxial cables in two steps: First we made open
and load calibrations by unwiring and wiring a 100 ohm chip resistor to the cables at the sample stage,
respectively. Then, the sample was connected to the cables and cooled to the lowest temperature of $\sim$ 1.2 K,
which is far below $T_c$ of $\alpha$-MoGe films and NbSe$_2$ crystals used in this study. Ensuring negligible
contact resistance and undetectable resistance of the sample, we made short calibration as function of frequency
as the second step.

During the ML measurement, we probed the dc voltage by ramping the dc-current up or down with a superimposed ac
current on top. The dc current value was obtained by measuring the dc voltage over a standard resistor. Other dc
transport measurements of resistive transition and dc current-voltage characteristics without ac current were
performed by using a dc current source (Keithley 220) and nanovoltmeters with proper filtering.

\section*{3. Results and Discussions}

Figure 1 displays a plot of the critical current $I_c$ vs field measured in 1.96 K for MG$\sharp$1. Here $I_c$
is determined from dc current-voltage ($I-V$) characteristics by a voltage criterion of 1 $\mu$V. As observed,
on increasing magnetic field $I_c$ starts to increase around 5.5 T (the onset field) and exhibits a broad peak
around 7 T (= $\mu_0H_p$). Above 7 T, it drops rapidly and vanishes around 7.8 T. We also plot the dynamic
critical current $I_{c,dyn}$, obtained by extrapolating the flux-flow behavior linearly to the zero voltage (for
definition, see the inset to Fig. 1). This characterizes the pinning strength (or dynamic friction) for driven
vortices in the flux-flow state. As observed, $I_{c,dyn}$ exhibits a pronounced peak against field: It starts to
show a rapid increase around 7 T and obtains a sharp maximum at about 7.6 T ($= \mu_0H_{p,dyn}$). Interestingly,
the peak field (onset field) of the dynamic critical current does not coincide with the corresponding peak
(onset) field for the critical current. The former appears in the higher part of the peak effect regime of the
latter, i.e., $H_{p,dyn} > H_p$. Such peak anomaly of the dynamic critical current is also observed in NbSe$_2$
crystals. In the conventional picture, the peak in the critical current marks the pinning induced, structural
transformation from weakly disordered to amorphous arrangement of vortices
\cite{WordenweberPRB1986}\cite{TroyanovskiPRL}\cite{Corbino}. Thus, from a phenomenological point of view,
another structural transformation in the flux-flow state may occur at the peak in the dynamic critical current.

\subsection*{Mode Locking}

In order to explore the dynamic structural change in the flux-flow state, we have employed the ML technique. To
show typical ML resonant features, we first present ML results below the peak effect regime. Shown in Fig. 2 is
a series of $I-V$ curves of MG$\sharp$2 measured with superimposed 10 MHz ac current of various amplitudes at
4.2 K and a field of 2.0 T ($< \mu_0H_p(4.2 \textrm{K})=$ 3.0 T). A curve denoted as DC represents the pure dc
$I-V$ curve without ac current $I_{ac}$. The application of $I_{ac}$ induces clear ML steps at equidistant
voltages denoted by $p/q$=1/1 and 2/1. The first step corresponds to the fundamental ($p/q=$1/1), and the other
is a higher harmonic ($p/q$=2/1). Also even higher harmonics are observed at higher voltages, but are not shown.
Because of the ML resonant condition of $v=(p/q)fa$ with integers $p, q$ and the lattice spacing of vortex
lattice $a$, these ML voltages are directly proportional to the resonant frequency of ac current according to
$V_{p/q}=vBl=(p/q)\Phi_0fl/a_\bot$ with the row spacing of vortex lattice $a_\bot(=\Phi_0/aB)$ and the vortex
density $B$ \cite{KKBPRL2005}. Step width at the fundamental (and also harmonic) ML resonance varies with
$I_{ac}$, exhibiting a oscillatory behavior like squared Bessel function of the first kind of the first (higher)
order  \cite{KKBPRB2004}. Since the oscillatory period depends on frequency, temperature, and magnetic field, we
use for consistency in the following analysis the maximum width $\Delta I_{max}$ of the fundamental resonance as
the amplitude of the ML step width.

In Fig. 3(a) we show how $\Delta I_{max}$ varies with frequency. At lower frequencies ($<$ 10 MHz), the ML step
width increases linearly with $f$, while at higher frequencies ($>$ 10 MHz) it levels off and becomes
independent of $f$. Such behavior is reasonably described by an empirical function of $\Delta I_{max} = I_s
\textrm{tanh}(f/0.7f_p)$, displayed by the solid curve \cite{KKBPRB2004}. From the fit we obtain a pinning
frequency $f_p$= 13 MHz and a saturated current width $I_s = 16 \mu A$ \cite{PinningFrequency}. However, the ML
width at lower frequencies deviates downward from the empirical function.  A linear extrapolation of the data
indicated by the dotted line seems to reveal a threshold behavior of the ML width, i.e., there is a finite
frequency (velocity) above which the ML resonance can be observed.

Such threshold behavior is pronounced in the peak effect regime, as can be seen in Fig. 3 (b), where the
frequency dependence is shown of the ML width measured at 4.2 K and magnetic fields of 3.25 T, 3.4 T, and 3.45 T
(above $\mu_0H_p= 3.0$ T). For instance, the linear extrapolation at 3.25 T clearly displays the threshold
behavior at an onset frequency $f_c \approx$ 1.4 MHz denoted by an arrow. Below $f_c$, no ML resonance feature
is detectable at any of the ac current amplitudes, indicating the absence of shear rigidity in the driven vortex
matter. Thus, $f_c$ marks a dynamic ordering frequency (or velocity) \cite{RutPRL2003}\cite{KKBPRB2004}. When
field is increased by 0.15 T to 3.4 T, $f_c$ increases by 1.7 MHz to $\approx$ 3.1 MHz. With further increase of
field by only 0.05 T to 3.45 T, $f_c$ shows a rapid increase by $\approx$ 2.2 MHz and becomes $\approx$ 5.3 MHz.
In slightly higher fields an even more rapid increase of $f_c$ is observed.


We summarize results of the onset frequency at 4.2 K in a semi-logarithmic plot of $f_c$ vs. magnetic field,
displayed in Fig. 4(b). In low fields, $f_c$ (depicted by solid diamonds) increases slowly with field. Around
the peak field of the critical current (=3.0 T), it turns to a rapid increase, followed by a small knee around
3.2 T. After the knee, it exhibits a diverging behavior towards just above 3.5 T. This is also displayed in a
linear plot given in the inset of Fig. 4(b). The open symbols represent the onset frequency determined from the
magnetic field dependence of the ML width \cite{KKBPRL2005}. Both data sets show the same diverging behavior of
$f_c (H)$.

The divergence of $f_c$ is in qualitative agreement with the dynamic ordering picture proposed by Koshelev and
Vinokur \cite{KosheleVinokurPRL1994}. The dynamic crystallization frequency can be given by
\cite{RutPRL2003}\cite{KKBPRB2004}
\begin{eqnarray}
f_c = \frac{v_c}{a} = \sqrt{\frac{3}{2\pi}}\cdot\frac{\gamma_u\rho_f}{\Phi_0^2 a^2 d k_B
T_M}\cdot\frac{1}{(1-T/T_M)}
 \label{fcTrelation}
\end{eqnarray}
with a disorder parameter $\gamma_u$ and the flux-flow resistivity $\rho_f$. This expression qualitatively
explains the divergence of $f_c$ (or $v_c$) against temperature observed in an $\alpha$-MoGe film
\cite{KKBAIP2006} and also in mesoscopic flow channels \cite{RutPRL2003}. Moreover, as displayed by the solid
curve in Fig. 4(b) (also in the inset), we find that the diverging behavior of $f_c$ against magnetic field is
approximated well by a similar diverging function of $f_c=f_0/(1-H/H_M)$ with $\mu_0H_M=3.53$ T (indicated by
the broken vertical line) and $f_0 = 0.11$ MHz over more than one decade in frequency range. These qualitative
agreements allow us to identify the field (or temperature) of the divergence as the thermal melting point of the
coherently flowing vortex lattice \cite{KKBPRL2005}. Away from $H_M$ at lower fields, the situation is
different. Here the onset of ML resonance occurs at small frequencies (velocities) and the influence of quenched
disorder dominates the ordering transition. The contribution of quenched disorder on $f_c$ may be determined
from following arguments: The disorder parameter is given by the amplitude of the pinning correlation function
of the random pinning potential which decays over the pinning radius $r_p$ \cite{KosheleVinokurPRL1994}. Using
the maximum pinning force $F_p$ per pinning site, we may write
\begin{eqnarray}
 \gamma_u=(N_p F_p^2(a/2\pi)^2/2)(r_p)^2 \approx Wa^6d/16\pi^2
 \label{disorderparameter}
\end{eqnarray}
where $N_p(=a^2dn_p)$ is the number of pinning sites per vortex line volume, $n_p$ is the density of pinning
centers, $W(=n_pF_p^2/2)$ is the Larkin-Ovchinnikov pinning parameter \cite{LarkinOch1979} and $r_p \approx a/2$
\cite{KesPRB1983}. Assuming $T_M-T \approx \mu_0(H_M-H)/S$, we find
\begin{eqnarray}
 f_c \approx f_0 \cdot\frac{1}{(1-H/H_M)}
 \label{fcHrelation}
\end{eqnarray}
with
\begin{eqnarray}
f_0 = \frac{1}{4\pi^2\sqrt{6\pi}}\cdot\frac{S \rho_f(H)
W(H)}{\mu_0H_Mk_BB^2}. \label{f0relation}
\end{eqnarray}
We know that the pinning properties for $\alpha$-MoGe films are well described by the delta-$T_c$ pinning, i.e.,
$W=Cb(1-b)^2$ with a reduced field $b(\equiv H/H_{c2}(T))$
\cite{KesPRB1983}\cite{WordenweberLTP1988}\cite{WordenweberPRB1986}. The flux flow resistivity for the amorphous
films in the GL regime can be written as $(\rho_f/\rho_n)^{-1}= \tilde{F}(b)/\sqrt{1-t}+1$ with a reduced
temperature $t\equiv T/T_c$ \cite{BerghuisPRB1993}. In fields measured here, i.e., $b=\mu_0H/\mu_0H_{c2} \geq$
1.4 T/4.54 T, the function $\tilde{F}(b)$ can be approximated by an interpolation formula
$\tilde{F}(b)=(1-b)^{3/2}(0.43+0.69(1-b))/b$ \cite{LarkinOvc}. Substituting these formula in Eq.
(\ref{f0relation}), together with $\mu_0H = B$, we find that $f_c$ varies with magnetic field as
\begin{eqnarray}
f_c \propto
\frac{(1-b)^2}{b(1+\tilde{F}(b)/\sqrt{1-t})}\cdot\frac{1}{1-H/H_M}.
\label{fcHrelation2}
\end{eqnarray}
As displayed by the dotted curve, this shows a slow increase with field away from $H \approx H_M$ and explains
our results below $H_p$. The proportionality constant extracted from the fit provides a rough estimate for the
coefficient $C$ in the expression for the pinning strength $W$, namely $C \approx 1.2*10^{-8}$N$^2/$m$^3$ at $T
\approx$ 0.7 $T_c$. This value is reasonably consistent with values obtained for previous transport measurements
on similar amorphous superconducting films \cite{KesPRB1983}\cite{WordenweberLTP1988}\cite{WordenweberPRB1986}.
Just above $H_p$, $f_c$ switches to a rapid increase up to the knee. This behavior follows neither the diverging
behavior close to $H_M$ displayed by a solid curve nor the low field behavior by a dotted curve. It indicates a
rapid enhancement of the disorder contribution (probably the pinning parameter $W$). Since the ordering
frequency (velocity) around $H_p$ is small and the disorder effect does not diminish significantly, this jump
behavior could be related to the underlying mechanism for the peak effect, i.e., the enhancement of the pinning
force caused by structural disordering transformation of static vortices at $H_p$
\cite{WordenweberPRB1986}\cite{TroyanovskiPRL} \cite{Corbino}.

We have systematically made ML and dc transport measurements in different temperatures on both
$\alpha$-MoGe films and NbSe$_2$ crystals. Results commonly show that $f_c$ exhibits the jump and
subsequent knee behavior around $H_p$.  The diverging behavior of $f_c$ occurs always in the higher part
of the peak effect regime and the divergent point lies in between the peak field and the second critical
field, i.e. $H_p < H_M < H_{c2}$. Thus, observed behavior of $f_c$ allows us to separate the thermal
melting marked by the divergent point from the disorder induced transformation marked by $H_p$.

\subsection*{Phase diagrams}

Based on these findings, let us map out a field-temperature phase diagram for $\alpha$-MoGe films of MG$\sharp$1
and MG$\sharp$2. To show results in two films together in Fig. 5(a), magnetic field and temperature are
normalized by the GL second critical field $H_{c2}^{GL}(0)(\equiv ST_c)$ and $T_c$, respectively.  For clarity,
results of MG$\sharp$1 and MG$\sharp$2 are depicted by black and red colored symbols, respectively. Good
agreement is visible in temperature dependencies of $h_{c2}(\equiv H_{c2}/H_{c2}^{GL}(0))$ plotted for the films
with solid red and black circles: On lowering temperature from $t=1$ $h_{c2} (t)$ increases linearly with $1-t$
displayed by a broken line. For $t \lesssim 0.8$, $h_{c2} (t)$ deviates downward from the linear dependence.
Such behavior is well described by the mean field curve (displayed by the solid curve) for the dirty limit in
Werthamer-Helfand-Hohenberg (WHH) theory \cite{Werthamer}, which is given by
\begin{equation}
\textrm{In}(t)=\Psi(1/2)-\Psi[1/2+(2/\pi^2)h_{c2}(t)/t] \label{Digamma}
\end{equation}
where $\Psi$ is the digamma function.

In Fig. 5(a) the results of $h_M(t)(\equiv H_M/H_{c2}^{GL}(0))$ for the films are also depicted by solid
squares. Focusing on results of MG$\sharp$1, we find that on lowering temperature from $t=1$ $h_M(t)$ increases
slowly and field distance between $h_M(t)$ and $h_{c2}(t)$ becomes larger. With further lowering $t$, $h_M(t)$
gradually increases faster and it becomes nearly parallel to $h_{c2}(t)$ for $t < 0.8$. Thus, the field distance
between $h_M(t)$ and $h_{c2}(t)$ is nearly constant for $t < 0.8$. Results of MG$\sharp$2 follow the $h_M(t)$
curve for MG$\sharp$1 well. Because of the thermal melting of the flowing vortex lattice marked by $h_M(t)$, the
regime in the phase diagram between $h_M(t)$ and $h_{c2}(t)$ should represent a flowing liquid of driven
vortices.

We also map out the field-temperature phase diagram for NbSe$_2$ crystals of NS$\sharp$1 and NS$\sharp$2 in Fig.
5(b). $h_{c2}(t)$ increases linearly with $1-t$ down to the lowest temperature measured. This is consistent with
thermodynamic studies on NbSe$_2$ crystals \cite{Toyota1976} \cite{Kobayashi1977}. In contrast to the results on
$\alpha$-MoGe films, $h_M (t)$ lies just below the mean field line except for low temperatures ($t\lesssim$
0.4). For instance, the field distance between $h_M (t)$ and $h_{c2} (t)$ at $t \sim$ 0.7 is 0.02 for NbSe$_2$
crystals, which is smaller than 0.08 for $\alpha$-MoGe films. In other words, the vortex liquid flow regime for
NbSe$_2$ crystals is much narrower than that for $\alpha$-MoGe films.

This difference should be related to the strength of thermal fluctuations, which may be characterized by the
Ginzburg number given by
\begin{equation}
Gi=\frac{1}{2}\cdot\left(\frac{k_BT_c\gamma}{4\pi\mu_0H_c(0)^2\xi_{ab}^3(0)}\right)^2 \label{Gi}
\end{equation}
where $\gamma$ is an anisotropy parameter, $H_c(0)$ is the thermodynamic critical field at $T=0$, and $\xi_{ab}
(0)$ is the coherence length for the ab plane at $T=0$ \cite{Blatter}. For estimation of $Gi$ on $\alpha$-MoGe
films, we use the dirty-limit expression for $\mu_0H_c(0)=ST_c/2.54\kappa$ with
$\kappa=3.54*10^4[\rho_nS]^{1/2}$ \cite{KesPRB1983} and $\gamma=1$. For NbSe$_2$ crystals, we use values of the
GL thermodynamic field $\mu_0H_c(0)=0.23$ T and $\gamma=$ 3, reported in the literature
\cite{Toyota1976}\cite{Kobayashi1977}. As listed in Table 1, $\alpha$-MoGe films have a nearly identical value
of $Gi \approx 8*10^{-5}$ which is two orders of magnitude larger than that for NbSe$_2$ crystals. This is
qualitatively consistent with the broader regime of the vortex liquid flow for $\alpha$-MoGe films.

Quantitative estimation of the distance between the thermodynamic melting curve and the mean field line can be
made by using the recent quantitative GL result proposed by Li and Rosenstein (LR) \cite{LiRosenPRB2002}. In
their theory, the thermodynamic melting transition sets in when the LLL-scaled temperature given by
\begin{equation}
a_T=-a_h\left(\sqrt{\frac{N_{Gi}}{2}}\pi t h\right)^{-2/3} \label{LLLtemp}
\end{equation}
is equal to -9.5. Here $N_{Gi}(=Gi/\pi^2)$ is their Ginzburg number and $a_h(=1-t-h)$ is the distance from the
mean field line. Substituting $Gi$ for $\alpha$-MoGe films and NbSe$_2$ crystals from table 1, we calculated how
far the melting curve is located from the mean field line ($1-t-h_{c2}(t)=0$). A solid blue curve in Fig. 5(a)
represents the calculated melting curve for MG$\sharp$1. Good agreement with $h_M(t)$ data is visible at high
temperatures ($t\gtrsim$ 0.6). We note that the melting curve at low temperatures ($t<$ 0.6) is not shown since
clear deviations of $h_{c2}$ from the linear dependence, which is presumed by the model, appear at low
temperatures. We also display the calculated melting curve for NS$\sharp$1 in Fig. 5(b). Because of small $Gi$,
the calculated melting curve lies just below the $h_{c2} (t)$ line. Again, good agreement of $h_M(t)$ with the
melting curve is visible at high temperatures ($t\geq 0.6$). Taking into account that this model is valid in
temperatures not far from $t=1$, the agreements of $h_M(t)$ with the LR model for both MG$\sharp$1 and
NS$\sharp$1 having more than two orders of magnitude difference in $Gi$ are quite remarkable, providing a
further link between the thermal melting point of the flowing lattice and the thermodynamic one of the static
and ideal vortex lattice. We note that the influence of applied currents upon the thermodynamic properties is
negligibly small \cite{current}.

We also plot the peak fields $h_p(\equiv H_p/H_{c2}^{GL}(0))$ of the critical current (open circles) in Fig.
5(a) and (b). In both cases they lie below the thermodynamic melting curve (or $h_M (t)$). This indicates that
for static vortices the pinning induced, structural transformation into amorphous array of vortices occurs prior
to the thermodynamic melting and therefore a crossover (or a continuous transition) from glassy to liquid states
may occur around the thermodynamic melting point. We emphasize that in the higher part of the peak effect regime
the dynamic melting features discussed above is observable only when vortices are driven fast enough and that
the elasticity is recovered by the dynamical ordering from the amorphous static state.

Finally we comment on the peak behavior in the dynamic critical current. We find that all the peak
points of the dynamic critical current observed for both $\alpha$-MoGe films and NbSe$_2$ crystals
coincide nearly with corresponding divergent points of $h_M(t)$ (not shown for clarity). This agreement
is exemplified in Figs. 4(a) and (b), where the peak point of the dynamic critical current occurs very
close to the divergence of $f_c$. These coincidences suggest that the peak point of the dynamic critical
current could be a good indication for the thermal melting of the coherent lattice flow.
However, it is not clear to us why the lower part of the peak anomaly observed in both $\alpha$-MoGe films and
NbSe$_2$ crystals change gradually and do not show a jump at the thermal (or thermodynamic) melting point. This
gradual change is possibly be related to an extrinsic origin like inhomogeneity in samples and/or geometrical
influences of edge \cite{EdgeXiao} and surface roughness \cite{SurfacePautrat}.

\section*{Summary}

In summary, employing the mode locking technique, we have presented the dynamic ordering of driven vortex matter
in the peak effect regime of both amorphous MoGe films and NbSe$_2$ crystals. The dynamic ordering frequency
(velocity) marked by the onset of the ML resonance exhibits a sharp increase and subsequent knee behavior around
the peak field of the critical current, followed by a striking diverging behavior in the higher part of the peak
effect regime. The former jump behavior indicates the rapid enhancement of pinning contribution, probably
originating from the underlying structural transformation of the static vortex lattice due to quenched disorder
at $H_p$. Meanwhile the latter diverging behavior is in good agreement with the dynamic ordering model and the
divergence point is identified as the thermal melting point of the driven vortex lattice.

The thermal melting point measured at each temperature lies in between the peak field and the second critical
field, and it coincides nearly with the peak point of the dynamic critical current. Mapping out the
field-temperature phase diagrams for amorphous MoGe films and NbSe$_2$ crystals, we find that the thermal
melting points follow well the thermodynamic melting curve for the static and ideal vortex lattice over wide
field/temperature ranges. These findings clearly separate the peak effect from thermal melting and provide a
link between static and dynamic melting phenomena.

\section*{Acknowledgements}
N. K. thanks E. Zeldov for useful comments on electronic contacts for NbSe$_2$ crystals. N. K thanks also S.
Bhattacharya, B. Rosenstein, G. Bel and V. M. Vinokur for useful comments. N. K used facilities in cryogenic
center in university of Tsukuba. This work was partly supported by the grant in Aid for Scientific research from
MEXT (the Ministry of Education, Culture, Sports, Science and Technology), and by the 21st Century COE (Center
of Excellence) Program, "Promotion of Creative Interdisciplinary Materials Science for Novel Functions" under
MEXT, Japan.

\end{multicols}

\newpage
\section*{Figure caption}
Fig. 1 Magnetic field dependence at 1.96 K of critical current $I_c$ (solid circles) and dynamic critical
current $I_{c,dyn}$ (open circles) for an $\alpha$-MoGe film (MG$\sharp$1). In the inset an $I-V$ curve measured
at 7.2 T is given. The solid line represents the linear extrapolation of the flux flow behavior to the zero
voltage. As indicated by an arrow, the intersectional point defines $I_{c,dyn}$.

Fig. 2  A set of dc current-voltage curves for an $\alpha$-MoGe film (MG$\sharp$2) measured at 2.0 T in 4.2 K
with superimposed ac currents (amplitude 3.3, 2.6, 2.1, 1.6, 1.3, 1.0, 0.82, 0.65, 0.52, 0.41, 0.33, and 0.26 mA
from left to right). A curve denoted as DC indicates the pure dc $I-V$ curve. Application of ac current induces
clear mode-locking current steps at equidistant voltages denoted by $p/q=$1/1 and 1/2.

Fig. 3  Frequency dependence of maximum current width $\Delta I_{max}$ of the fundamental ML step for an
$\alpha$-MoGe film (MG$\sharp$2) measured (a) at 2.0 T, far below the peak effect regime, and (b) at fields of
3.25 T, 3.40 T and 3.45 T in the peak effect regime in 4.2 K. A solid curve represents an empirical function for
the ML width given in text \cite{KKBPRB2004}. Dotted lines represent the linear low frequency behavior of the ML
width. Arrows mark the onset frequency $f_c$ for ML resonance.


Fig. 4  Magnetic field dependence of(a) critical current $I_c$, dynamic critical current $I_{c,dyn}$ and (b) the
onset frequency $f_c$ for an $\alpha$-MoGe film (MG$\sharp$2) at 4.2 K. Peak points of the critical current and
the dynamic critical current are indicated by arrows in (a). The inset to (b) displays the plot of $f_c$ on a
linear frequency scale.

Fig. 5. (Color online) Magnetic field-temperature phase diagrams for (a) $\alpha$-MoGe films and (b) NbSe$_2$
crystals. The magnetic field and temperature are normalized by superconducting transition temperature $T_c$ and
the Ginzburg-Landau second critical field $H_{c2}^{GL}(0)(\equiv ST_c)$. Plotted are reduced second critical
fields $h_{c2}$ (small solid circles), divergence fields $h_M$ of the ordering frequency (solid squares) and
peak fields $h_p$ of the critical current (open circles). In (a), black and red colored symbols correspond to
results for MG$\sharp$1 and MG$\sharp$2, respectively. In (b), black (red) colored symbols are for NS$\sharp$1
(NS$\sharp$2). In (a) a solid black curve represents the mean field curve for the dirty-limit in WHH theory
\cite{Werthamer}. For comparison, the linear dependence of $h_{c2}$ on $1-t$ is displayed by a broken line in
(a) (also in (b)).  A dotted curve in (a) is guide to the eye. Blue curves in (a) and (b) represent
thermodynamic melting curves obtained from a quantitative GL model \cite{LiRosenPRB2002} for MG$\sharp$1 and
NS$\sharp$1 crystals, respectively.


Table 1.  Properties of amorphous MoGe films and NbSe$_2$ crystals.










\newpage
\begin{table}[h]
\caption{Properties of amorphous MoGe films and NbSe$_2$ crystals}
 \label{Table1}
  \begin{center}
   \begin{tabular}{ccccccccc} Material&Sample&$T_c$(K)&$\rho_n$ (8
K)($\mu\Omega$m)&$-\mu_0dH_{c2}/dT|_{T_c}$(T/K)&$l$(mm)&$w$(mm) &$d(\mu$m)&$G_i$\\ \hline
MoGe&MG$\sharp$1&6.11&1.84&2.6&1.2&0.3&0.33&8.9*10$^{-5}$\\
 &MG$\sharp$2&6.05&1.72&2.7&1.2& 0.3&0.33&8.1*10$^{-5}$\\
NbSe$_2$&NS$\sharp$1&7.1&0.040&0.75&0.8&0.4&1.8\tablenotemark[1]&6.1*10$^{-7}$\\
 &NS$\sharp$2&7.2&0.034&0.72&0.72&0.63&0.9\tablenotemark[1]&5.8*10$^{-7}$\\
\end{tabular}
\tablenotetext[1]{The thickness was estimated from the normal resistance at the room temperature $R(295
\textrm{K})$ by $d=(l/w) \rho_n (295 \textrm{K})/R(295 \textrm{K})$ with $\rho_n (295 \textrm{K})=1.2\mu\Omega$m
\cite{KKBPRL2005}.}
\end{center}
\end{table}


\begin{thebibliography}{00}
\bibitem{PinningFrequency} The pinning frequency is much lager than $f_p=J_c\rho_f/aB \approx$ 3MHz (8.2MHz) estimated from
the (dynamic) critical current, flux flow resistivity $\rho_f$ and $B=\mu_0H$. This differene would be explained
by possible distribution in the pinning frequency due to plasticity in driven vortices.
\bibitem{current} The influence of the applied currents upon thermodynamic properties for the isotropic superconductor
may be characterized by the dimensionless velocity $\tilde{v} = 6\pi\mu_0V\xi^3\kappa^2/l\Phi_0\rho_nh(1-t)$
\cite{BelPreprint}. Substituting material parameters and the ML voltage for 10 MHz by the thermal melting point,
we find $\tilde{v} \approx 4*10^{-5}$ at $h_M (0.7) \approx$ 0.22 for MG$\sharp$2 and $\tilde{v} \approx
3*10^{-4}$ at $h_M (0.7) \approx$ 0.28 for NS$\sharp$1. These are negligibly small compared to the distance from
the mean field line $a_h (t=0.7) \approx$ 0.08 and 0.02 for MG$\sharp$2 and NS$\sharp$1, respectively.
\end{thebibliography}
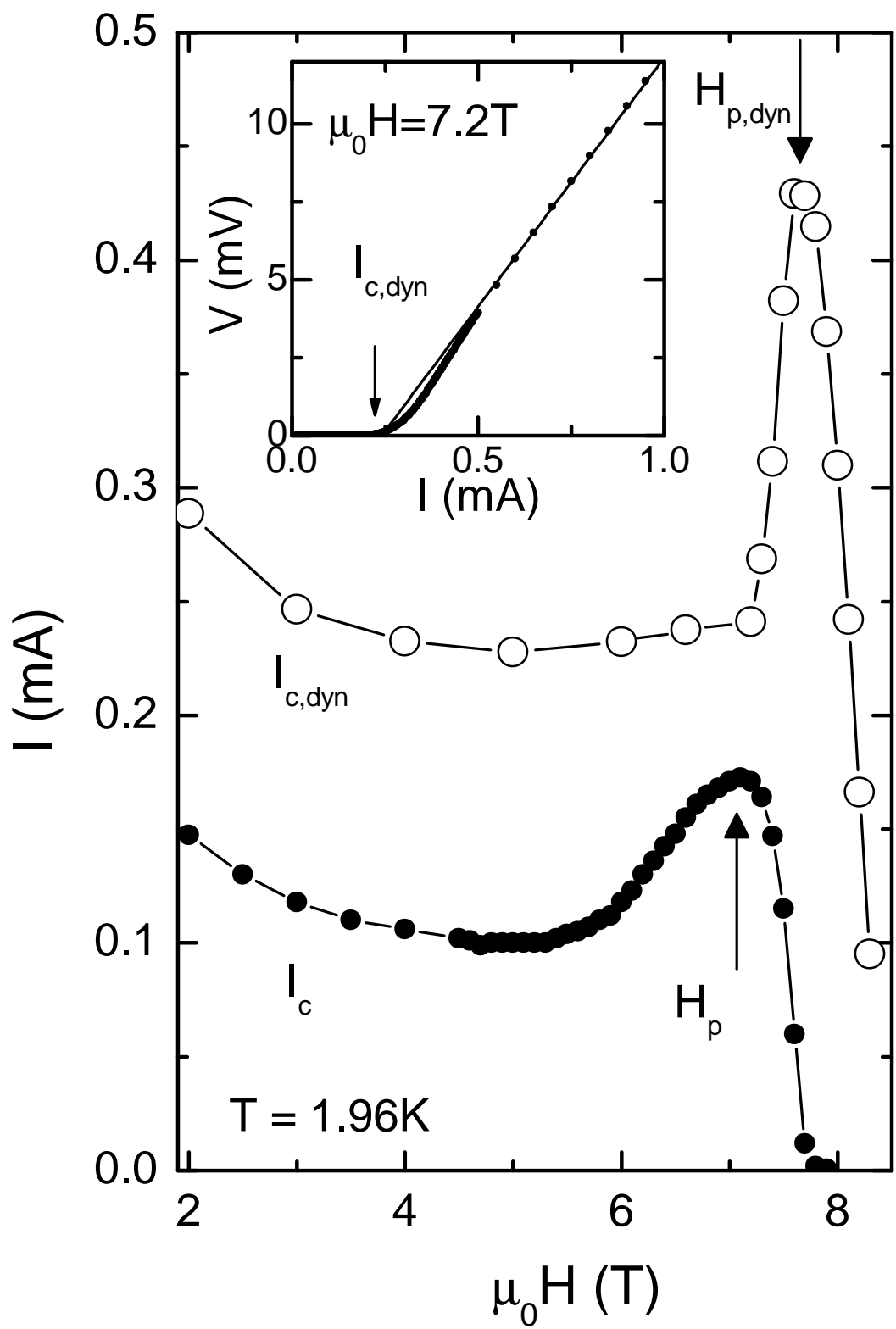
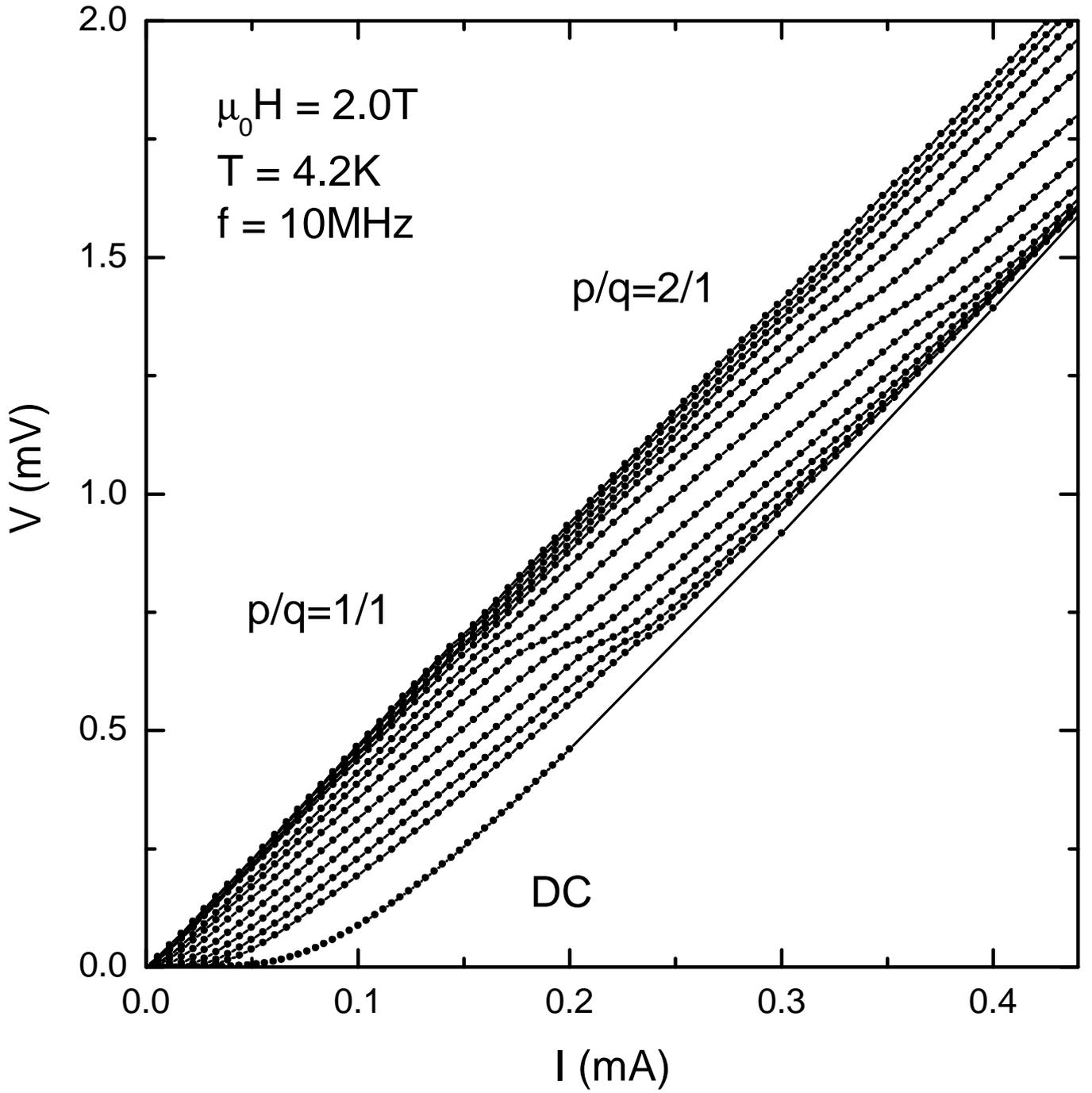
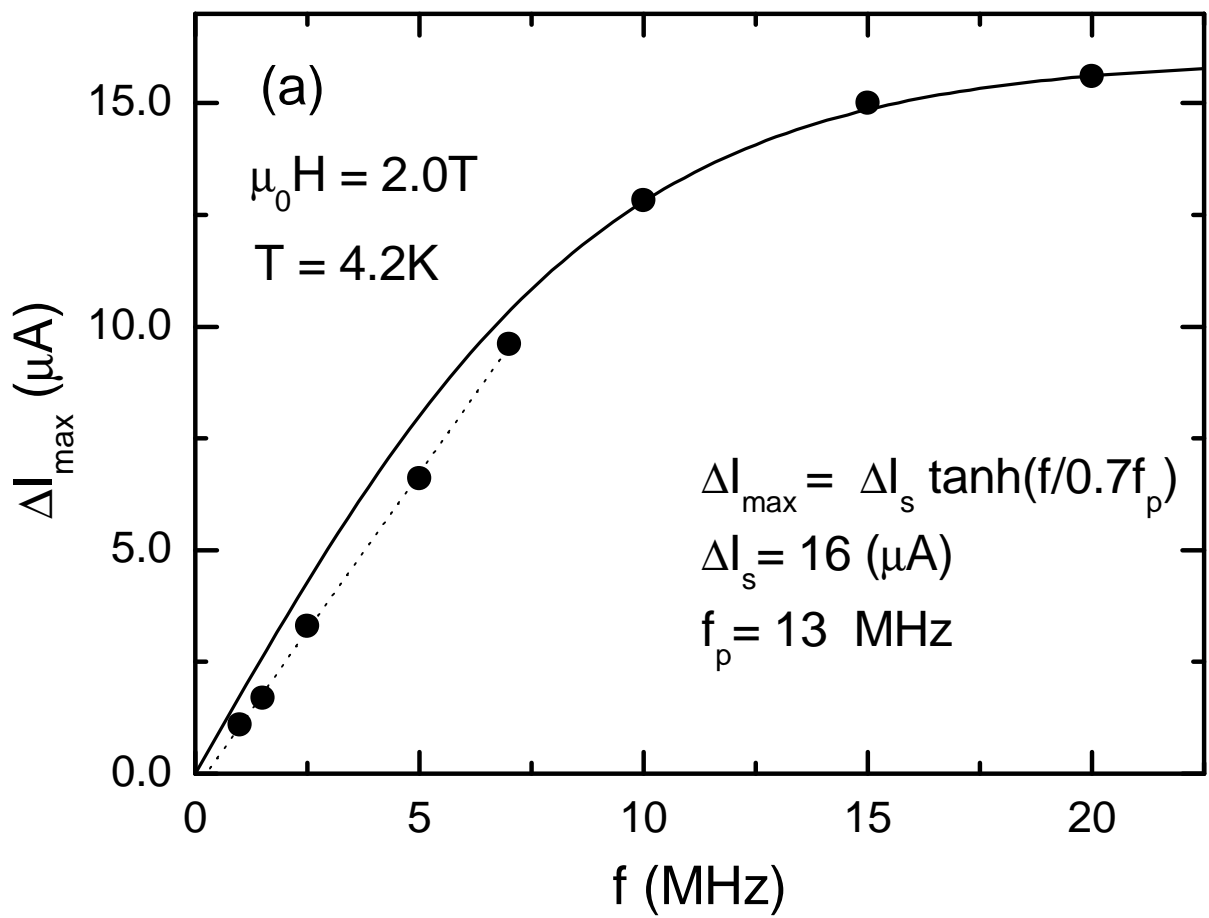
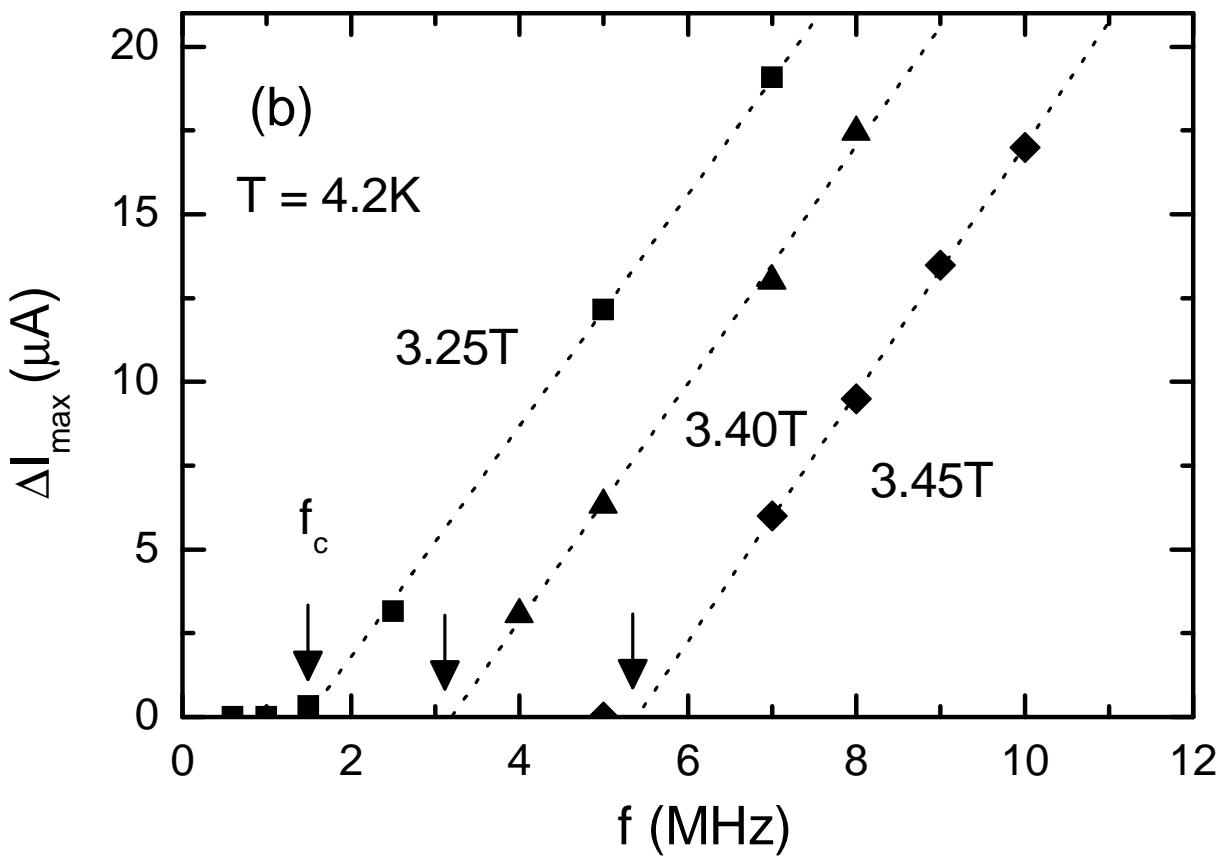
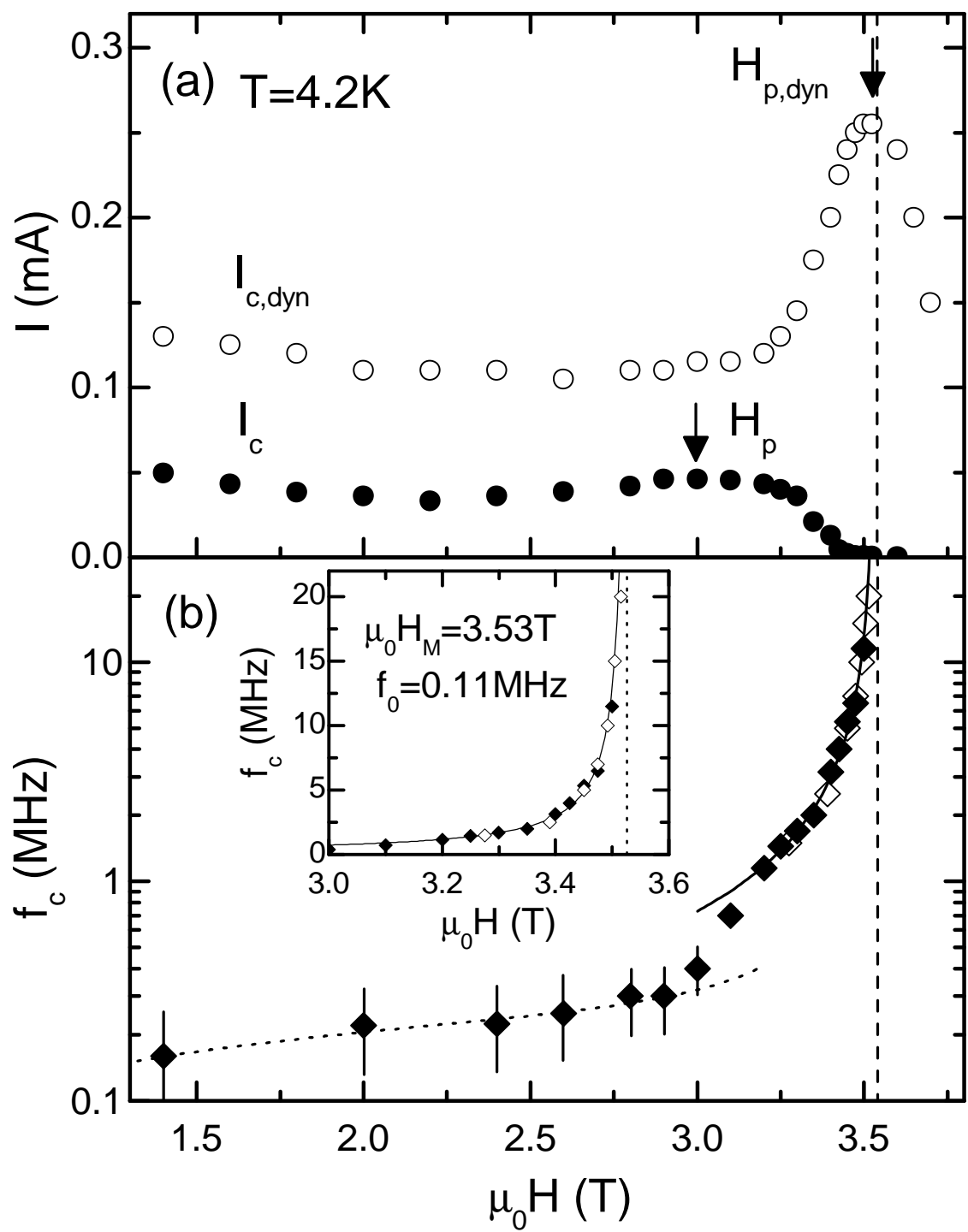
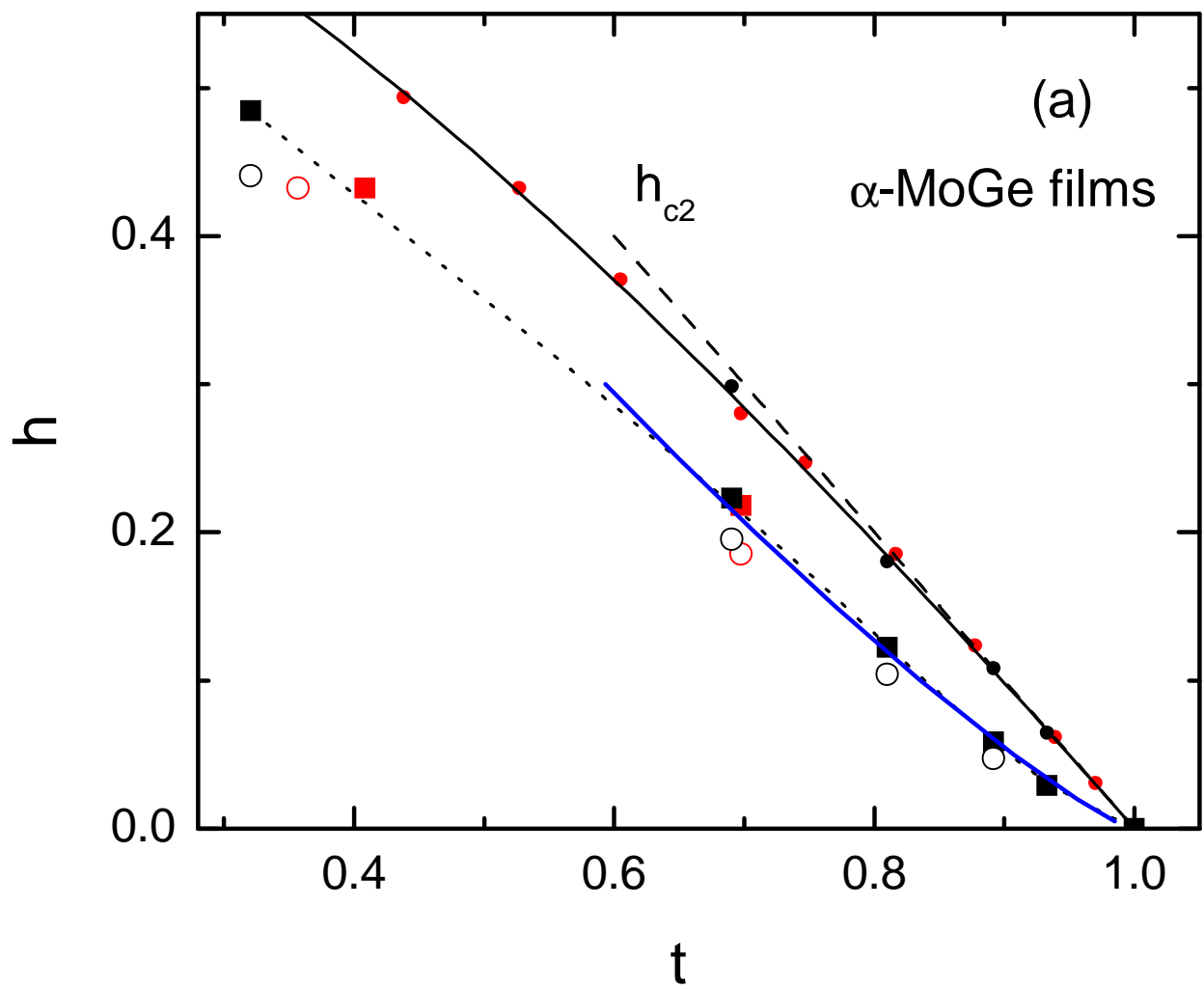
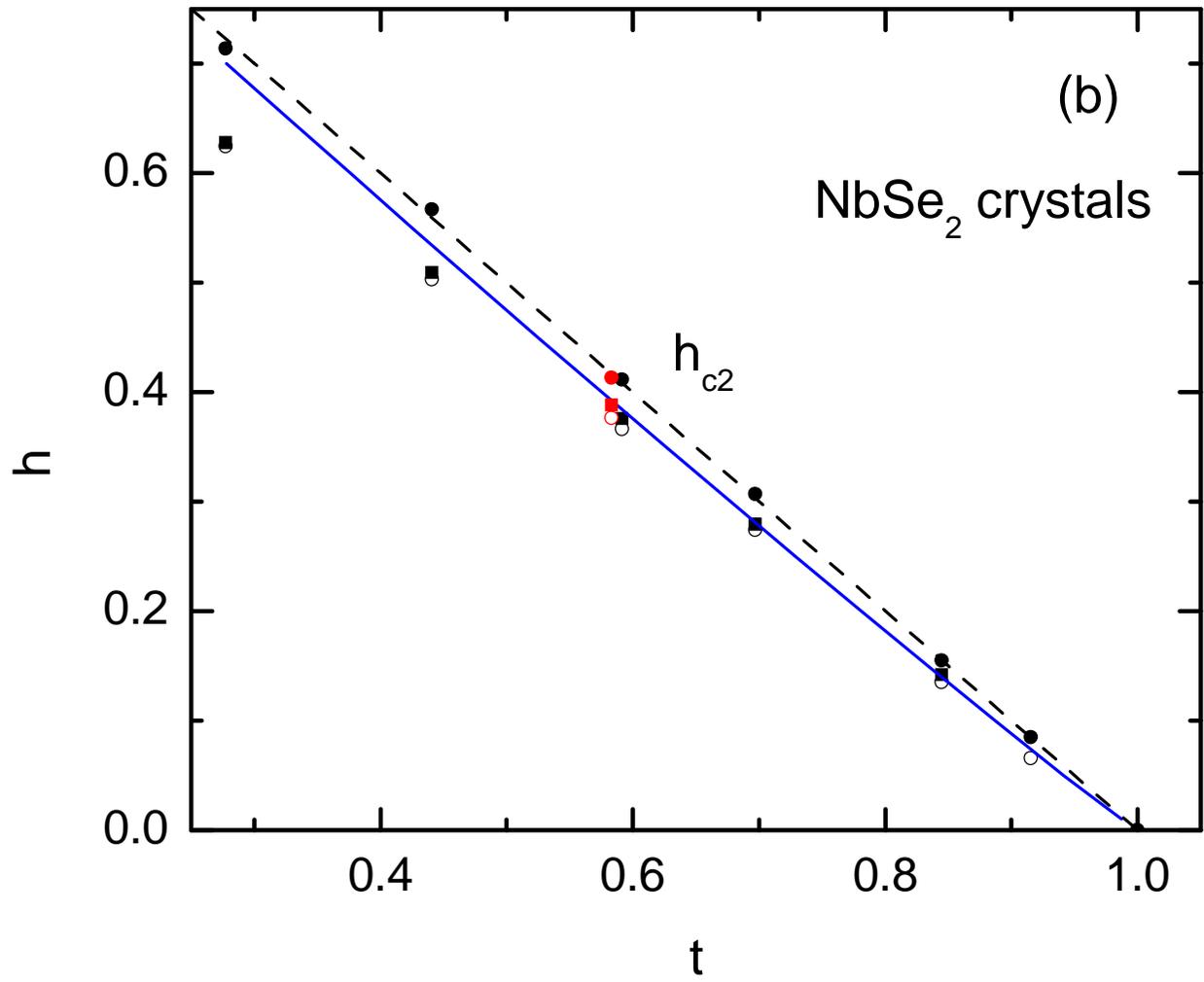
\end{document}